\documentclass[10pt]{article}

\usepackage{amsfonts,latexsym,amssymb}

\def\geq{\geqslant}

\title
{
The study of distributed computing algorithms
 by multithread applications
}
\author{Ahmet A. Husainov}
\date{}

\begin{document}
\maketitle
\begin{abstract}
The material in this note is used as an introduction to distributed
algorithms in a four year course on software and automatic control
system in the computer technology department of the Komsomolsk-on-Amur
state technical university. All our the program examples are
written in Borland C/C++  5.02 for Windows 95/98/2000/NT/XP, and
hence suit to compile and execute by Visual C/C++.
% and Borland C++ Builder.
We consider the following approaches of the distributed computing:

 The conversion of recursive algorithms to multithread applications

 A realization of the pairing algorithm

 The building of wave systems by Petri nets and object oriented programming.

\end{abstract}

2000 Mathematics Subject Classification 68M14, 68Q85.


\smallskip

%\section*{Introduction}
      
\section{Threads and the conversion of recursive programs }

A thread is declared by its descriptor. We create the thread by the
API function
\begin{verbatim}     
      HANDLE h=CreateThread (NULL, 0, Thr, (void *)p, 0, 0);
\end{verbatim}     
which loads the subroutine 
\begin{verbatim}
DWORD WINAPI Thr (void *p)
{
         ...               
}
\end{verbatim}
and returns the descriptor h of the thread if the creation
is successful and h=NULL otherwise.
The argument p points to a list of parameters of the function Thr().
The subroutine call
\begin{verbatim}
      WaitForSingleObject(h, INFINITE);
\end{verbatim}
leads to the waiting of the thread termination.

If a program has a recursive function,
then we can convert all the function calls to the thread creations.
This conversion leads to a multithread application.
We consider the Hoare quick sort for example:

\begin{verbatim}
#include <stdlib.h>
#include <time.h>
#include <iostream.h>
#define N 100
int x[N];
void q_sort(int l, int u)
{
int i, j, m, t; int temp;
 if(l < u)
 {
     t = x[l]; m = l;
     for (i = l + 1; i <= u; i++)
     if(x[i] < t)
     {
     	m++; temp = x[m]; x[m] = x[i]; x[i] = temp;
     }
     temp = x[l]; x[l] = x[m]; x[m] = temp;
     q_sort(l, m - 1); q_sort(m + 1, u);
 }
}
main()
{
     int i; randomize();
     for (i = 0; i < N; i++) x[i] = random(100);
     q_sort(0, N - 1);
     for (i = 0; i < N; i++) cout << " " << x[i];
}
\end{verbatim}     
The subroutine q\_sort() has two parameters. Hence we must define a structure

\begin{verbatim}
struct arg
{
     int left, right;
};
\end{verbatim}     
for the parameters passing. We obtain the following text after the conversion:
\begin{verbatim}     
#include <windows.h>
#include <stdlib.h>
#include <time.h>
#include <iostream.h>
#include <conio.h>
#define N 100
struct arg
{
  int left, right;
};
int x[N];
DWORD WINAPI q_sort(void* p)
{
int l=((arg *)p)->left, u=((arg *)p)->right;
int i,j,m,t,temp;
  if (l<u)
  {
      t = x[l]; m = l;
      for (i=l+1; i<=u; i++)
      if (x[i]<t)
      {
          m++; temp=x[m]; x[m]=x[i]; x[i]=temp;
      }
      temp = x[l]; x[l]= x[m]; x[m]=temp;
      arg *r1 = new arg, *r2 = new arg;
      HANDLE H[2];
      r1->left=l; r1->right= m-1;
      r2->left= m+1; r2->right = u;
      H[0]= CreateThread(0,0,q_sort, (void *)r1,0,0);
      H[1]= CreateThread(0,0,q_sort, (void *)r2,0,0);
      for (i=0; i<2; i++)
          WaitForSingleObject(H[i],INFINITE);
      delete r1; delete r2;
  }
  return 1;
}
void main()
{
int i;
  randomize();
  for (i=0; i<N; i++) x[i]=random(100);
  for (i=0; i<N; i++) cout << " "<< x[i];
  cout << "\n";
  arg a;
  a.left = 0; a.right = N-1;
  q_sort(&a);
  for (i=0; i<N; i++) cout << " "<< x[i];
  getch();
}
\end{verbatim}     
The first call of the recursive function remains
in the main program whereas the calls in q\_sort() convert to CreateThread().
A project for this program must include the options
''console application" and the ''multithread application".
Hence the creation of the project is given by the menu command
$File\rightarrow New\rightarrow Project$ with Target Model ''Console"
and the option ''Multithread". The files
''.def" and ''.rc" must be deleted from the project.

     The program puts a sequence of 100 random numbers and
then these numbers are displayed in the undecreasing order.
     
\section{The pairing algorithm implementation}

It is possible the implementation of a multiply applied associative operation
by the recursive subroutines. The subroutine has two parameters
with numbers of first and last elements of an array:
\begin{verbatim}    
int sum(int l, int r) // x[l] + ... + x[r]
{
  if(l == r) return x[l];
  else return sum(l, (l+r+1)/2 - 1) + sum((l+r+1)/2, r);
}
\end{verbatim}   
This subroutine is called in the main program by s=sum(0,n-1).
Hence, we obtain
by the conversion to the multithread application the following text:
\begin{verbatim}          
#include <windows.h>
#include <stdlib.h>
#include <time.h>
#include <iostream.h>
#include <conio.h>
#define N 32
struct arg
{
  int l, r, rez;
};
int x[N];
DWORD WINAPI sum(void* s)
{
int i, l=((arg *)s)->l, r=((arg *)s)->r;
  if (l==r) ((arg *)s)->rez = x[l];
  else
  {
    arg *r1 = new arg, *r2 = new arg;
    HANDLE H[2];
      r1->l=l; r1->r= (l+r+1)/2-1;
      r2->l=(l+r+1)/2; r2->r = r;
      H[0]= CreateThread(0,0,sum, (void *)r1,0,0);
      H[1]= CreateThread(0,0,sum, (void *)r2,0,0);
      for (i=0; i<2; i++)
        WaitForSingleObject(H[i],INFINITE);
      ((arg *)s)->rez = (r1->rez)+(r2->rez);
      delete r1; delete r2;
  }
  return 1;
}
int sum0()
{
int s=0, i;
  for (i=0; i<N; i++) s+=x[i];
  return s;
}
void main()
{
int i;
  randomize();
  for (i=0; i<N; i++) x[i]=random(100);
  arg t;
  t.l = 0; t.r = N-1;
  sum(&t);
  cout << "\n sum obtained = "<< t.rez;
  cout << "\n sum is = "<< sum0();
  getch();
}
\end{verbatim}     
     
     Consider the pairing algorithm for the computing of the polynomial
     $p(x)= a_0+a_1{x}+\cdots+a_n{x^n}$
     values by Horner's scheme:
$$
p_0= a_n; \quad p_1=xp_0+a_{n-1}; \quad p_2= xp_1+ a_{n-2}; \quad
\cdots \quad p_n= xp_{n-1}+a_0\,.
$$
with $p(x)=p_n$. If it possible to write $p_k=xp_{k-1}+a_{n-k}$
as the equation
$$
\left(
\begin{array}{c}
p_k\\
1\\
\end{array}
\right)
=
\left(
\begin{array}{cc}
x & a_{n-k}\\
0 & 1\\
\end{array}
\right)
\left(
\begin{array}{c}
p_{k-1}\\
1\\
\end{array}
\right)\,.
$$
Therefore
$$
\left(
\begin{array}{c}
p(x)\\
1\\
\end{array}
\right)
=
\left(
\begin{array}{cc}
x & a_0\\
0 & 1\\
\end{array}
\right)
\left(
\begin{array}{cc}
x & a_1\\
0 & 1\\
\end{array}
\right)
\cdots
\left(
\begin{array}{cc}
x & a_{n-1}\\
0 & 1\\
\end{array}
\right)
\left(
\begin{array}{c}
a_n\\
1\\
\end{array}
\right)\,,
$$
and we can to compute the values of p(x) by the pairing algorithm
for the matrix multiplications.

Now we write the recursive subroutine of the matrix multiplication.
The subroutine get parameters contained in a structure
\begin{verbatim}
struct arg
{
  int l, r;
  double p, q; // Result
};
\end{verbatim}
The data are read from the external array a[i] and the variable $x$.
We have a  text of the subroutine:
\begin{verbatim}
void prod(void *s)
{
  int i, l = ((arg*)s) -> l, r = ((arg*)s) -> r;
  if(l == r)
  {
     ((arg*)s) -> p = x;
     ((arg*)s) -> q = x[l] ;
  }      
  else
  {
     arg *r1 = new arg, *r2 = new arg;
     r1 -> l = l; r1 -> r1 = (l + r + 1)/2 - 1;
     r2 -> l = (l + r + 1)/2; r2 -> r = r;
     ((arg )s) -> p = (r1 -> p)(r2 -> p);             
     ((arg )s) -> q = (r1 -> p)(r2 -> q) + (r1 -> q); 
     delete r1; delete r2;
  }
}
\end{verbatim}
Converting this subroutine we obtain the following multithread
application where the values 
$a_0+a_1 x+a_2 x^2+ \cdots +a_N x^N$ are calculated
by the pairing of the Horner scheme
\begin{verbatim}
#include <windows.h>      // multithread opportunity
#include <stdlib.h>       // random numbers generation
#include <time.h>         // the first random number
#include <iostream.h>     // input/output
#include <conio.h>        // for the getch()
#include <dos.h>          // debugging by sleep()
#define N 20              // degree of the polynomial
double x;                 // the value of the variable
struct arg                // for the parameter settings
{
  int l, r;               // left and right numbers of 2õ2-marix
  double p, q;            // the first matrix string entries
};
int a[N+1];               // coefficients of the polynomial
DWORD WINAPI prod(void* s)// the thread subroutine
{
int i, l=((arg *)s) -> l, r=((arg *)s) -> r;
  sleep(random(2));       // the delay for the debugging
  if (l == r)
  {
      ((arg *)s) -> p = x;    // the coefficient
      ((arg *)s) ->q = a[r];  // settings
  }
  else
  {
      arg *r1 = new arg, *r2 = new arg;
        HANDLE H[2];                             // descriptors
        r1 -> l = l; r1 -> r = (l + r + 1)/2 - 1;// boundaries
        r2 -> l = (l + r + 1)/2; r2 -> r = r;         
      H[0] = CreateThread(0,0,prod, (void *)r1, 0,0);// create
      H[1] = CreateThread(0,0,prod, (void *)r2,0,0); // threads
        for (i = 0; i < 2; i++)                      // wait of
        WaitForSingleObject(H[i],INFINITE);      // terminates
      	((arg *)s) -> p = (r1 -> p)*(r2 -> p);
      	((arg *)s) -> q = (r1 -> p)*(r2 -> q)+(r1 -> q);
      	delete r1; delete r2;
  }
       return 1;
}
void main()
{
int i,j;
double w=0, v, z;
  randomize();
  for (i=0; i <=N; i++)
        a[i] = random(10);      
  arg t;                      // parameter structure
  for (x = 0; x < 10; x += 1) // values of x
  {
        t.l = 0; t.r = N;     // boundaries 
        prod(&t);             // call of the subroutine
        cout << "\n Polynomial value = " << t.q; // output
        z = 0;
      for (i = 0; i <= N; i++)// checking cycle
      {
        v = 1;
        for (j = 0; j < i; j++) v = v*x;   // computation of xi
          z += v*a[i];        // the polynomial values
      }
      cout << " == "<< z;     // displays for the checking
      getch();                // wait of the keyword press
  }
}
\end{verbatim}
     
In fact the pairing algorithm computes in this application
the product of $N+1$ matricies. The first string entries
of the product will be equal to $x^{N+1}$  and $p*a[N]+q$,
where $p$  and $q$ are the first string entries of the product of
$N$ matrices. Thus the program displays the values  $p(x) = p*a[N]+q$,
and values obtained by the single thread program for the checking:
\begin{verbatim}
      Polynomial value = 9 == 9
      Polynomial value = 53 == 53
      Polynomial value = 1.99675e+06  == 1.99675e+06
      Polynomial value =  4.74122e+09 == 4.74122e+09
      Polynomial value = 1.31457e+12  == 1.31457e+12
      Polynomial value = 1.07123e+14  == 1.07123e+14
      Polynomial value = 3.96656e+15  == 3.96656e+15
      Polynomial value = 8.47447e+16  == 8.47447e+16
      Polynomial value = 1.20753e+18  == 1.20753e+18
      Polynomial value = 1.26117e+19  == 1.26117e+19
\end{verbatim}
     
\section{The synchronizing mechanism}

Suppose that a few threads work with one variable in the global memory.
It is possible the access of one at most
a thread to this variable in any time.
The access synchronization of these threads is called 
{\em a serialization problem}. A solution of the serialization problem is
based on a semaphore mechanism.
{\em A semaphore} is the data structure which consists of one
integer number $s \geq 0$ and two functions P(s) and V(s)
which are possible to unite in the following class:
\begin{verbatim}     
class Semaphore
{
      volative int s;                // a semaphore counter
  public:
      Semaphore(int init): s(init){} // initial counter value
      void P() {while (s==0); s--;}  // wait and locks
      void V() {s++;}                // signal operation
}
\end{verbatim}
Suppose that the functions of the semaphore class are realized
by the hardware.

Semaphores allow us solve the serialization problem. For example,
consider the concurrent computing of the vector $(x_1,x_2, \cdots, x_n)$
length square. We load $n$ threads for the independent calculations
of $x_i*x_i$. Every thread locks the global variable $s$ after
the calculation of $d = x_i*x_i$ and yields the action $s = s + d$.

     We obtain the following program scheme:
\begin{verbatim}
double x[n];    // vector coordinates
Semaphore s(1); // initialize s = 1
double d = 0;   // the global variable
main ()
{
      /* here the threads are loaded ... */
}
\end{verbatim}
     
     The thread subroutine can be contains the following operators:
\begin{verbatim}   
double y = x[i]*x[i];
s.P(); // the lockout of  d resource
d = d + y;
s.V(); // the signal operation
\end{verbatim}
     
     Our aim is to write this program in the Borland C/C++. We will be
use the following subroutines for the semaphore operations. The part of
the constructor is belong to the function:
\begin{verbatim}    
      HANDLE s = CreateSemaphore (NULL, i, n, NULL);
\end{verbatim}
where i is the initial value of the semaphore counter whereas $n$ is
the maximal value. The role of the call s.P() is played by
\begin{verbatim}     
      WaitForSingleObject(s, INFINITE) ;
\end{verbatim}
and there call s.V() is executed by
\begin{verbatim}
      ReleaseSemaphore(s, 1, NULL).
\end{verbatim}   

These subroutines allow us to write the following solution of
the serialization problem:
\begin{verbatim}
#include <windows.h>
#include <stdlib.h>
#include <time.h>
#include <iostream.h>
#include <conio.h>
#define n 100
volatile int j=0;
HANDLE mut;
double d=0;
double x[n];
DWORD WINAPI Thr(LPVOID v)
{
  double  *w = (double *)v; double u=(*w)*(*w);
  WaitForSingleObject(mut, INFINITE);
          d= d+u; j++;
  ReleaseSemaphore(mut,1, NULL);
  return 1;
}
double sum0()
{
double s0=0; int k;
  for (k=0; k<n; k++)
        s0+= x[k]*x[k];
  return s0;
}
void main()
{
  int i;
  randomize(); 
  for (i=0; i<n; i++) x[i]=(0.+random(100))/100+5;
  mut = CreateSemaphore(NULL,1,1, NULL);
  for (i=0; i<n; i++)
  {
        CreateThread(NULL,0,Thr, (void *)(&x[i]),0,0);
  }
  while (j<n);
  cout << "\nValue obtained by threads = "<< d;
  cout << "\nValue computed by function = "<< sum0();
  getch();
}
\end{verbatim}
     
\section{A channel class for the producer and consumer problem }
     
     Using the idea the Occam we define a class which objects are channels.
Each channel is a data queue organized as an array.
Suppose that this array contains the double precision numbers.
The access to the queue is provided by the overload operations
$<<$ for the writing and $>>$ for the reading. Using a classical
solution we define the class of the channel. We write this definition
into the file channel.h:
\begin{verbatim}
// channel.h
template<class Type>
class channel
{
  Type *buf;                // the buffer for a queue
  int size;                 // the buffer size
  HANDLE s,                 // the access semaphore
    empty,                  // number of free places in the queue 
    full;                   // number of full places in the queue
    int countr, countw;     // read/write counters
  public:
  channel (int n): size(n)  // constructor
  {
    buf = new Type [n];     // the memory of the buffer
    countr=0; countw=0;
    s=CreateSemaphore(NULL,1,1,NULL); // semaphore initialization
    empty=CreateSemaphore(NULL,n,n,NULL); // n free places
    full=CreateSemaphore(NULL,0,n,NULL);  // no full places
  }
  void operator << (Type d)               // write to channel
  {
    WaitForSingleObject(empty, INFINITE); // wait free places
    WaitForSingleObject(s, INFINITE);     // take buffer 
    buf[countw++]=d;                      // write to queue
    if (countw==size) countw=0; 
    ReleaseSemaphore(s,1,NULL);               
    ReleaseSemaphore(full,1,NULL);        // full++
  }
  void operator >> (Type& d)    // read from channel
  {
    WaitForSingleObject(full, INFINITE); // wait of data
    WaitForSingleObject(s, INFINITE);    // take buffer
    d = buf[countr]; countr++;           // read
    if (countr==size) countr=0; 
    ReleaseSemaphore(s,1,NULL);          
    ReleaseSemaphore(empty,1,NULL);      // empty++
  }
};
\end{verbatim}
It is a generalization with the template type instead of double.
Now we have the following solution of the producer and consumer problem:
     
\begin{verbatim}
#include <windows.h>
#include <iostream.h>
#include "channel.h"
double out[20];
DWORD WINAPI producer(LPVOID v) 
{
int j;
double d1;
  for (j=0; j<20; j++)
  {
          d1 = (double)(1+j);
          *(channel<double>*)v<<d1;
  }
  return 1;
}
DWORD WINAPI consumer(LPVOID v) 
{
  int k;
  for (k=0; k<20; k++)
  {
          *(channel<double>*)v>>out[k];
  }
  return 1;
}
void main()
{
channel<double> c0(10);
int i=0;
  CreateThread(NULL,0,producer, (LPVOID) &c0,0,0);
  CreateThread(NULL,0,consumer, (LPVOID) &c0,0,0);
  cout<<"\n"; for(i=0;i<20;i++) cout << " "<< out[i];
  getch();
  cout<<"\n"; for(i=0;i<20;i++) cout << " "<< out[i];
}
\end{verbatim}    

Here the producer writes to the channel the sequence of numbers.
The consumer reads this sequence and writes it to the array out[].
The application displays 20 values of 0, and the sequence $1$, $2$,
$\cdots$, $20$ after the key input.

\section{Wave systems and their Petri nets}
     
 A wave system is the generalization of the pipeline.
Consider the computing of the values $y_n=f(g(x_n))$ with
$n=0, 1, 2, \cdots$. It is possible to load the threads which
states are described by the following Petri net:

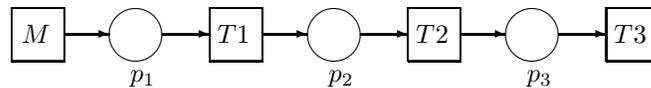
\begin{center}
\begin{picture}(280,40)
\multiput(57,20)(75,0){3}
{\circle{20}}

\put(55,2){$p_1$}
\put(130,2){$p_2$}
\put(205,2){$p_3$}

\multiput(10,10)(75,0){4}
{\line(1,0){20}}
\multiput(10,10)(75,0){4}
{\line(0,1){20}}
\multiput(30,30)(75,0){4}
{\line(-1,0){20}}
\multiput(30,30)(75,0){4}
{\line(0,-1){20}}

\multiput(30,20)(75,0){3}
{\vector(1,0){17}}
\multiput(67,20)(75,0){3}
{\vector(1,0){18}}

\put(14,17){$M$}
\put(88,17){$T1$}
\put(163,17){$T2$}
\put(238,17){$T3$}
\end{picture}
\end{center}
where the main program $M$ and threads $T1$, $T2$, $T3$ are executed
concurrently. The main program generated random numbers $x_n$
and writes $x_n$  to the channel corresponding to the channel $p_1$.
The thread $T1$ computes $g(x_n)$ and writes it to the channel
$p_2$. The thread $T2$ computes $f(g(x_n))$ and the thread
$T3$ gets the data from the channel $p_3$ and puts to the display.
Such a system is called the pipeline and has the following generalization.
For example, consider the computation of values
$$
        x_n= u_n+sin(v_n^2), \quad y_n = exp(sin(u_n-v_n)).
$$
Applicate the flow control with the channel communications.
The factorization of these operations leads us to the 10 threads
executed concurrently.
The threads correspond to the transitions of the Petri net which is
shown in Figure \ref{ff}.
\unitlength=0.9pt
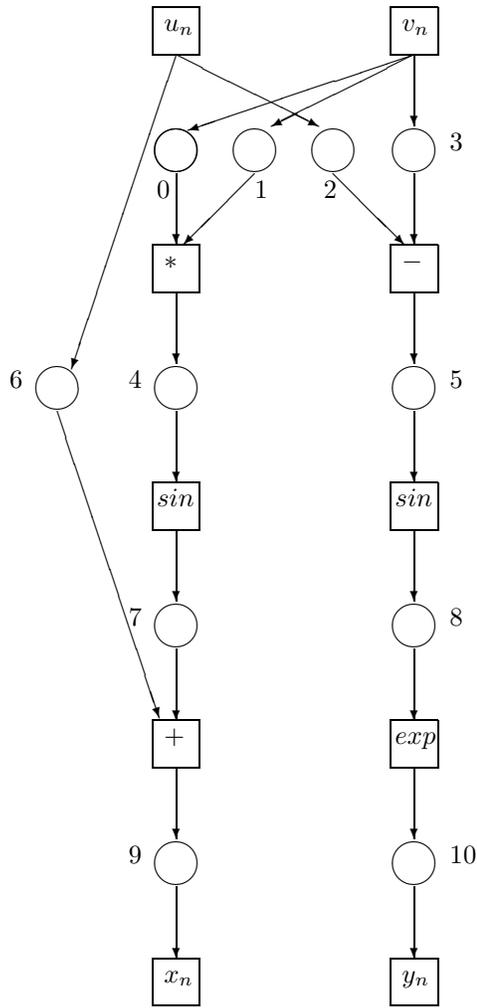
\begin{figure}
\begin{center}
\begin{picture}(300,500)
\multiput(100,100)(0,100){4}
{\circle{20}}
\multiput(200,100)(0,100){4}
{\circle{20}}
\multiput(100,400)(33,0){3}
{\circle{20}}
\multiput(90,40)(0,100){5}
{\line(1,0){20}}
\multiput(90,40)(0,100){5}
{\line(0,1){20}}
\multiput(110,60)(0,100){5}
{\line(-1,0){20}}
\multiput(110,60)(0,100){5}
{\line(0,-1){20}}
\multiput(190,40)(0,100){5}
{\line(1,0){20}}
\multiput(190,40)(0,100){5}
{\line(0,1){20}}
\multiput(210,60)(0,100){5}
{\line(-1,0){20}}
\multiput(210,60)(0,100){5}
{\line(0,-1){20}}
\put(50,300)
{\circle{20}}

\multiput(100,90)(0,100){4}
{\vector(0,-1){30}}
\multiput(200,90)(0,100){4}
{\vector(0,-1){30}}

\multiput(100,140)(0,100){3}
{\vector(0,-1){30}}
\multiput(200,140)(0,100){4}
{\vector(0,-1){30}}

\put(50,290)
{\vector(1,-3){43}}
\put(100,440)
{\vector(-1,-3){44}}
\put(100,440)
{\vector(2,-1){60}}
\put(200,440)
{\vector(-2,-1){60}}
\put(200,440)
{\vector(-3,-1){95}}
\put(133,390)
{\vector(-1,-1){30}}
\put(166,390)
{\vector(1,-1){30}}

\put(95,50){$x_n$}
\put(195,50){$y_n$}
\put(95,150){$+$}
\put(192,150){$exp$}
\put(92,250){$sin$}
\put(192,250){$sin$}
\put(95,350){$*$}
\put(195,350){$-$}
\put(95,450){$u_n$}
\put(195,450){$v_n$}
\put(80,100){$9$}
\put(215,100){$10$}
\put(80,200){$7$}
\put(215,200){$8$}
\put(80,300){$4$}
\put(215,300){$5$}
\put(30,300){$6$}
\put(92,380){$0$}
\put(215,400){$3$}
\put(133,380){$1$}
\put(162,380){$2$}

\end{picture}
\caption{The Petri net of the wave system}
\label{ff}
\end{center}
\end{figure}

     The Petri net consists of  11 places corresponding to the
channels and 10 transitions corresponding to the threads. 
Output of  $(x_n,y_n)$ will be yield in the main program.
Thus the computations will be yield in the seven threads. The array of
channels
\begin{verbatim}     
channel c[11]={5,5,5,5,5,5,5,5,5,5,5};
\end{verbatim}     
is declared in the main program. Each channel is a queue of
five elements of the array of double precision numbers.
These channels are connected with  global pointers declared as
\verb"channel *ps[11]". The threads execute concurrently.
Each of threads waits an input data and then computes an operation.
For example, the subroutine \verb"mult()" gets the data from the channels
\verb"c[0]" and \verb"c[1]" and sends their product to the channel \verb"c[4]".

\begin{verbatim}
#include <windows.h>
#include <stdlib.h>
#include <iostream.h>
#include <math.h>
#define n 15
class channel
{
  double *buf;
  int size;
  HANDLE s,
  empty,
  full;
  int countr, countw;
 public:
   channel(int nn):size(nn)
   {
      buf = new double[nn];
      countr=0; countw=0;
      s=CreateSemaphore(NULL,1,1,NULL);
      empty=CreateSemaphore(NULL,nn,nn,NULL);
      full= CreateSemaphore(NULL,0,nn,NULL);
   }
   void operator<<(double d)
   {
      WaitForSingleObject(empty, INFINITE);
      WaitForSingleObject(s, INFINITE);
      buf[countw++]=d;
      if (countw==size) countw=0;
      ReleaseSemaphore(s,1, NULL);
      ReleaseSemaphore(full,1, NULL);
   }
   void operator>>(double& d)
   {
      WaitForSingleObject(full, INFINITE);
      WaitForSingleObject(s, INFINITE);
      d = buf[countr]; countr++; 
      if (countr==size) countr=0;
      ReleaseSemaphore(s,1,NULL);
      ReleaseSemaphore(empty,1,NULL);
   }
};

channel *pc[11];

DWORD WINAPI mult(LPVOID)
{
   int j; double  d1,d2;
   for (j=0;j<n;j++)
   {
     *pc[0]>>d1; *pc[1]>>d2;
     *pc[4]<<(d1*d2);
   }
   return 1;
}
DWORD WINAPI minus(LPVOID)
{
   int j;
   double  d1,d2;
   for (j=0;j<n;j++)
   {
     *pc[2]>>d1; *pc[3]>>d2;
     *pc[5]<<(d1-d2);
   }
   return 1;
}
DWORD WINAPI sinus(LPVOID)
{
   int j;
   double  d1;
   for (j=0;j<n;j++)
   {
      *pc[4]>>d1;   	*pc[7]<<sin(d1);
   }
   return 1;
}
DWORD WINAPI sinus2(LPVOID)
{
   int j;
   double  d1;
   for (j=0;j<n;j++)
   {
      *pc[5]>>d1;   	*pc[8]<<sin(d1);
   }
	return 1;
}
DWORD WINAPI plus(LPVOID)
{
   int j;
   double  d1,d2;
   for (j=0;j<n;j++)
   {
      *pc[6]>>d1; *pc[7]>>d2;
      *pc[9]<<(d1+d2);
   }
   return 1;
}
DWORD WINAPI expo(LPVOID)
{
   int j;
   double  d1;
   for (j=0;j<n;j++)
   {
      *pc[8]>>d1;   	*pc[10]<<exp(d1);
   }
   return 1;
}
DWORD WINAPI input(LPVOID)
{
   int i; double u,v;
   for (i=0; i<n;i++)
   {
      u=i; v=i/2;
      *pc[0]<<v; *pc[1]<<v;
      *pc[2]<<u; *pc[3]<<v; *pc[6]<<u;
   }
   return 1;
}

void main()
{
   channel c[11]={5,5,5,5,5,5,5,5,5,5,5};
   int i; double x,y;
   for (i=0;i<11;i++) pc[i]=&c[i];
   CreateThread(NULL,0,mult, 0,0,0);
   CreateThread(NULL,0,minus, 0,0,0);
   CreateThread(NULL,0,sinus, 0,0,0);
   CreateThread(NULL,0,sinus2, 0,0,0);
   CreateThread(NULL,0,plus, 0,0,0);
   CreateThread(NULL,0,expo, 0,0,0);
   CreateThread(NULL,0,input, 0,0,0);
   for (i=0; i<n;i++)
   {  
      c[9]>>x; c[10]>>y;  cout<<"\n"<<x;
      cout<<"=="; cout<<(double)(i+sin(0.+(i/2)*(i/2)));
      cout<<" "<<y<<"=="<<(double)(exp(sin(i-i/2)));
   }
   cout<<"\n";
}
\end{verbatim} 
   
     The application displays:

\begin{verbatim}
0==0 1==1
1==1 2.31978==2.31978
2.84147==2.84147 2.31978==2.31978
3.84147==3.84147 2.48258==2.48258
3.2432==3.2432 2.48258==2.48258
4.2432==4.2432 1.15156==1.15156
6.41212==6.41212 1.15156==1.15156
7.41212==7.41212 0.469164==0.469164
7.7121==7.7121 0.469164==0.469164
8.7121==8.7121 0.383305==0.383305
9.86765==9.86765 0.383305==0.383305
10.8676==10.8676 0.756226==0.756226
11.0082==11.0082 0.756226==0.756226
12.0082==12.0082 1.92897==1.92897
13.0462==13.0462 1.92897==1.92897
\end{verbatim}

The comparison of the results shows that the wave system is well defined.
     
 We conclude that there is the well opportunity for 
the  study of the concurrent algorithms on the every personal computer.

\begin{flushleft}
Department of Computer Technologies,\\
Komsomolsk-on-Amur State Technical University,\\
prosp. Lenina, 27, Komsomolsk-on-Amur, Russia, 681013\\
Email: husainov@knastu.ru
\end{flushleft}

\end{document}